\documentclass[prl,showpacs,onecolumn,preprint]{revtex4}

\usepackage{amsmath}
\usepackage{amsfonts}

\begin{document}

\preprint{}
\title[Random Geometric Phase Sequence]{Random geometric phase sequence due
to topological effects in our brane world from extra dimensions}
\author{Jian Fu}
\affiliation{State Key Lab of Modern Optical Instrumentation, Department of Optical
Engineering, Zhejiang University, Hangzhou 310027, China}

\begin{abstract}
Using Kaluza-Klein theory we discuss the quantum mechanics of a particle in
the background of a domain wall (brane) embedded in extra dimensions. We
show that the geometric phases associated with the particle depend on the
topological features of those spacetimes. Using a cohomological modeling
schema, we deduce a random phase sequence composed of the geometric phases
accompanying the periodic evolution over the spacetimes. The random phase
sequence is demonstrated some properties that could be experimental
verification. We argue that it is related to the nonlocality of quantum
entanglement.
\end{abstract}

\date{today}
\startpage{1}
\email{jianfu@zju.edu.cn}
\maketitle

Quantum entanglement, one of the most fascinating and important features in
quantum theory, is widely appreciated as an essential ingredient in quantum
computations \cite{Nielsen,Bennett,Einstein,Jozsa}. It demonstrates one of
the most remarkable aspects of quantum theory is the incompatibility of
quantum nonlocality with local-realistic theories \cite{Bell}. Recently, an
effective simulation of quantum entanglement using classical fields
modulated with pseudorandom phase sequences is proposed \cite{Fu}. Inspired
by this, we argue that each quantum particle might be characterized by the
wavefunction with a unique random phase sequence, namely the quantum
particle might own some unknown intrinsic phase mechanism. The quantum
scenario demonstrates quantum nonlocality i.e. violation of Bell's
inequality, free from \textquotedblleft spooky action at a
distance\textquotedblright .

As an alternative theory adopting gravity into quantum theory, superstring
theory draws a lot of attention \cite{Green,Michio,Witten,Horava}. Different
from other theories, superstring theory has perfect mathematical and
physical structure, but hasn't been experimentally validated directly for
required unconceivable energy levels. Recently some researches have focused
non-perturbative effects of extra dimensions, hoping it can be observed
through the experiment. In Kaluza-Klein theory, the extra dimensions being
rolled up to Planck length makes domain wall (brane) surrounded by particle
string. This particle string winding movement around topological defects
will bring non-perturbative geometric effects \cite%
{Dvali,Furtado,Capozziello,Kikkawa}.

In quantum mechanics, the global phase of particle wavefunction has been
ignored, due to the Born rule probability interpretation of a state vector
at a single moment of time. The notion of a global geometric phase factor
has been introduced in quantum mechanics by Berry \cite{Berry,Shapere} and
formulated in fiber bundle theoretic terms by Simon \cite{Simon}. In 1959
Aharonov and Bohm showed that there exists the global geometric phase in the
interference pattern of two coherent wavefunctions of a charged particle due
to the existence of an electromagnetic field constrained within an
infinitely long solenoid even the strength vanished \cite{Aharonov}.
Recently, Zafiris showed that a sheaf cohomological understanding of the
origin of global geometric phase factors paves the way for understanding the
global symmetry group of a quantum spectral beam \cite{Zafiris}. They employ
the local phase invariance of a quantum state vector is related to the
observation of global gauge-invariant topological and geometric phase
factors accompanying the periodic evolution over a space of control
variables.

In this letter we will discuss the argument from two aspects: (1) what
physical mechanisms to generate random phase sequences for quantum
particles; (2) what verifiable theoretical predictions deduced from the
physical mechanisms. First we research geometric phases of particle string
wound in the brane of extra dimensions based on the Kaluza-Klein theory.
Then we research random phase sequences superposed of multiple geometric
phases in the control variable space $X$ of genus $g$ in the framework of
the cohomology theory proposed in Ref. \cite{Zafiris}. Finally, we deduce
three theoretical predictions that would be verified by future experiments.

In the conventional Kaluza-Klein approach the Universe has a topology $%
M_{4}\otimes K$ where $M_{4}$ is our four dimensional Minkowski space and $K$
is some compact manifold, with the volume typically set by a fundamental
Planck length $l_{p_{f}}=1/M_{P_{f}}$. Our world is a domain wall (brane)
embedded in the extra-dimensional space with a rolled up dimension with size 
$R\sim l_{p_{f}}$. In this physical picture, the rolled extra dimension has
one important difference from the particle theory version \cite{Green,Michio}%
. A closed string can get wound several times around a rolled up the
dimension. When a particle string does this, the particle string
oscillations have a winding mode. The winding modes add a symmetry to the
theory not present in particle physics.

Now we consider the geometric phase due to the winding modes \cite%
{Dvali,Furtado,Capozziello,Kikkawa}. The geometric phase is very similar to
the Aharonov-Bohm phase of a charged particle traversing a loop including a
magnetic flux \cite{Aharonov}, which is a nontrivial topological effect in a
multiply connected space. According to Ref. \cite{Dvali}, a toy model is to
assume that $K=S_{1}$ with nontrivial closed curve homotopy class $%
H_{1}\left( S_{1}\right) $. This has the isometry group $G=U\left( 1\right) $%
, broken by the brane down to identity. The brane position on $S_{1}$ is
parameterized by one scalar modulus $\sigma $. This position can slowly vary
with $x_{\mu }$, the four-dimensional space-time point. Thus, the
four-dimensional observer will perceive $\sigma \left( x_{\mu }\right) $ as
a low-energy scalar field on $M_{4}$. The particle string in the brane
vibrates and periodically moves surrounding the extra dimensions. In the toy
model, the target space of $\sigma $ is obviously a circle $S_{1}$, which
the corresponding fundamental group $H_{1}\left( S_{1}\right) =%
\mathbb{Z}
$. After a cyclic evolution of the environment parameters, the geometric
phase can be expressed abstractly%
\begin{equation}
\beta \left( C\right) =\oint\nolimits_{C}\left\langle \phi \left\vert
i\nabla \phi \right. \right\rangle =i\int_{0}^{\sigma \left( \tau \right)
=2\pi }\left\langle \phi \left( \sigma \right) \right\vert \frac{d}{d\sigma }%
\left\vert \phi \left( \sigma \right) \right\rangle d\sigma  \label{eq1}
\end{equation}%
where $\sigma $ and $\tau $ are coordinates on the brane representing space
and time along the string. The non-integrable geometric phase $\beta \left(
C\right) $ tells us something about the geometry of the circuit and about
regions of the brane characterizing (for example, enclosed by) the circuit.
Whereas $\beta \left( C\right) $ is the holonomy due to parallel transport
around a circuit, in analogy to the eletromagnetic Aharonov-Bohm effect \cite%
{Furtado}. The geometry of the brane is important in Berry's formulation
because geometric and topological features can present obstructions to a
global definition of the string phases \cite{Pedder}. This is why the
geometric phase is independent on the size of the extra-dimension even
rolled up in a circle of radius $R\sim l_{p_{f}}$. We note again that the
time dependence of geometric phase $\beta \left( C\right) $ is implicitly
introduced via time-parameterized paths on $K$, and $\beta \left( C\right)
=\beta \left( T\right) $ during the string vibrating within a finite
temporal period $T$. We obtain the geometric phase $\beta \left( \tau
+nT\right) =n\beta \left( T\right) \delta \left( \tau +nT\right) $, where $n$
is an integer, the winding number. In conclusion, the geometric phase must
be periodically appeared if the string vibrating without any disturbance.

Superstring theory dictates that the universe could be a topology $%
M_{4}\otimes K$ with extra space $K$ with multiple dimensions, which could
be a Calabi--Yau manifold. In Ref. \cite{Zafiris}, a sheaf-cohomological
concept is introduced to research the relation to the observation of global
gauge-invariant topological and geometric phase factors accompanying the
periodic evolution over a space $X$ of control variables. We consider the
parametric dependence of the particle string vibrating in the brane is
analogy to the control variable space $X$ and the notion of a vector sheaf
generalizes as Ref. \cite{Zafiris}. Considering the homology group $%
H_{1}\left( X\right) $ is composed of 1-dimensional homology class in $X$
space, the winding path $\sigma $ of the string in the brane is isomorphism
of homology cycle $\gamma \in H_{1}\left( X\right) $. According Ref. \cite%
{Zafiris}, a cohomology class in the cohomology group $H^{1}\left( X,U\left(
1\right) \right) $ can be evaluated at the homology cycle $\gamma $ by means
of the pairing:%
\begin{equation}
H_{1}\left( X\right) \times H^{1}\left( X,U\left( 1\right) \right)
\rightarrow U\left( 1\right)  \label{eq2}
\end{equation}%
to obtain a global observable gauge-invariant geometric phase factor $\beta
\in U\left( 1\right) $.

Further, we consider $X$ space is a compact Riemann surface of genus $g$.
Using simplicial complex analysis, we obtain a typical base $\left[ \gamma
_{1}\right] ,\left[ \gamma _{2}\right] ,\ldots \left[ \gamma _{2g}\right] $
of $H_{1}\left( X\right) $ composed of all homology classes and the
dimension $2g$ of cohomology group $H^{1}\left( X,U\left( 1\right) \right) $ 
\cite{Bredon}. We assume the winding path $\sigma $ of the particle string
is a simplicial 1-chain is a formal sum of 1-cycles: 
\begin{equation}
\sigma =\sum_{i=1}^{2g}m_{i}\gamma _{i}  \label{eq3}
\end{equation}%
where $m_{i}\in 
\mathbb{Z}
$. Based on the previous analysis, we assume that the string obtain
geometric phases $\beta _{i}$ in each homology cycle $\gamma _{i}\in
H_{1}\left( X\right) $. The total geometric phase can be express: 
\begin{equation}
\beta \left( \tau \right) =\sum_{i=1}^{2g}\beta _{i}\left( \tau
+m_{i}T_{i}\right) =\sum_{i=1}^{2g}m_{i}\beta _{i}\left( T_{i}\right) \delta
\left( \tau +m_{i}T_{i}\right)   \label{eq4}
\end{equation}%
where $m_{i}$ is the winding number. The geometric phases $\beta _{i}$ and
periods $T_{i}$ of the particle string sweeping each homology cycle $\gamma
_{i}$ should be different. According to the almost-periodic theory, the
total geometric phase is almost-periodic if any two periods of geometric
phases are incommensurable \cite{Besicovitch,Corduneanu}. As a result, the
total geometric phase becomes an almost-periodic random sequence \cite%
{Keh,Hong,Spitters}.

According the qualitative analysis of the superposed geometric phase, we
propose three theoretical predictions:

\textit{Superposition}: The geometric phase of quantum particle due to
topological defects of extra-dimensions is a phase sequence superposed of
multiple periodic functions, and the maximum count of the periodic functions
should be twice of genus $g$ of $X$ space.

\textit{Almost-periodic}: The geometric phase sequence is almost-periodic.

\textit{Random}: The geometric phase sequence is an almost-periodic random
sequence.

In Ref. \cite{Fu}, an effective simulation of quantum entanglement using
classical fields modulated with pseudorandom phase sequences is proposed.
Consider the simplest case that the mode $\left\vert 1\right\rangle $ of the
classical fields $\left\vert \psi _{a}\right\rangle $ and $\left\vert \psi
_{b}\right\rangle $ are exchanged by a mode exchanger constituted by mode
splitters and combiners as shown in Ref. \cite{Fu}. When the modes are
exchanged, the phase sequences modulated on classical fields also are
exchanged. The classical fields efficiently simulate two entangled
particles. However, it will be rather complicated for quantum particles,
which involves particle string interaction in the framework of superstring
theory \cite{Green,Michio}. Consider the simple case, the base of homology
group $H_{1}\left( X\right) $ and the winding modes of two particle strings
are exchanged through some string interactions. As a result, the random
geometric phase functions (sequences) $\beta ^{\left( a\right) }\left( \tau
\right) $ and $\beta ^{\left( b\right) }\left( \tau \right) $ of the
superposition states are also exchanged similar to the classical simulation: 
\begin{eqnarray}
\left\vert \psi _{a}^{^{\prime }}\right\rangle  &=&\frac{e^{i\beta ^{\left(
a\right) }\left( \tau \right) }}{\sqrt{2}}\left( \left\vert 0\right\rangle
_{a}+e^{i\gamma ^{\left( a\right) }\left( \tau \right) }\left\vert
1\right\rangle _{b}\right)   \label{eq5} \\
\left\vert \psi _{b}^{^{\prime }}\right\rangle  &=&\frac{e^{i\beta ^{\left(
b\right) }\left( \tau \right) }}{\sqrt{2}}\left( \left\vert 0\right\rangle
_{b}+e^{i\gamma ^{\left( b\right) }\left( \tau \right) }\left\vert
1\right\rangle _{a}\right)   \notag
\end{eqnarray}%
where $\gamma ^{\left( a\right) }\left( \tau \right) =-\gamma ^{\left(
b\right) }\left( \tau \right) =\beta ^{\left( b\right) }\left( \tau \right)
-\beta ^{\left( a\right) }\left( \tau \right) $ is also an almost-periodic
random function. We obtain the results of the particles in the measurement $%
P\left( \theta _{a},\tau \right) =\cos \left( \theta _{a},\gamma ^{\left(
a\right) }\left( \tau \right) \right) $ and $P\left( \theta _{b},\tau
\right) =\cos \left( \theta _{b},\gamma ^{\left( b\right) }\left( \tau
\right) \right) $. Then we define the temporal correlation function: 
\begin{equation}
E\left( \theta _{a},\theta _{b};t\right) =\frac{1}{C}\int_{0}^{t}P\left(
\theta _{a},\tau \right) P\left( \theta _{b},\tau \right) d\tau =\cos \left(
\theta _{a}+\theta _{b}\right) +\frac{1}{C}\int_{0}^{t}\cos \left( \theta
_{a}-\theta _{b}+2\gamma ^{\left( a\right) }\left( \tau \right) \right)
d\tau   \label{eq6}
\end{equation}%
where $C$ is the normalized coefficient. According to Riesc-Fischer theorem,
the correlation function $E\left( \theta _{a},\theta _{b};t\right) $ can be
related with the trigonometric polynomials and Fourier analysis \cite%
{Besicovitch}. Due to the almost-periodic random of $\gamma ^{\left(
a\right) }\left( \tau \right) $, we obtain 
\begin{equation}
\lim_{t\rightarrow \infty }\frac{1}{C}\int_{0}^{t}\cos \left( \theta
_{a}-\theta _{b}+2\gamma ^{\left( a\right) }\left( \tau \right) \right)
d\tau \rightarrow 0  \label{eq7}
\end{equation}%
and $E\left( \theta _{a},\theta _{b}\right) =\cos \left( \theta _{a}+\theta
_{b}\right) $ demostrates violation of Bell's inequality for two quantum
particles. The correlation function $E\left( \theta _{a},\theta
_{b};t\right) $ should be a useful tool to analysis the motion of particle
strings in the brane.

Long term since, quantum mechanics suffers from the complete and nonlocality
doubts. There might be two long-standing beliefs for this: (1) the
randomness of quantum only caused by the measurement process; (2) the
neglect of the overall phase of wavefunction due to no contribution to the
probability distribution. Based on the previous analysis, we propose a new
phase mechanism to explain the nonlocality of quantum mechanics. The
randomness of quantum might be originated from the random geometric phase
sequence of particle string wound in the brane of extra dimensions. The
random geometric phase sequence due to the topological features of $X$ space
might be not only confined between different particles also as a relative
phase sequence between different eigenstates of the measurement operators,
which also results in the randomness in quantum measurement of a single
particle. It is very important how to verify these predictions in the
experiment. We believe Bose-Einstein condensation \cite{Dalfovo} might be a
good experiment demonstration because the random geometric phase sequence
should be amplified to macroscopic quantum phenomena.

\quad


\begin{thebibliography}{99}
\bibitem{Nielsen} M. A. Nielsen and I. L. Chuang, \textit{Quantum
Computation and Quantum Information} (Cambridge University Press, Cambridge,
2000).

\bibitem{Einstein} A. Einstein, B. Podolsky, and N. Rosen, Phys. Rev. 
\textbf{47}, 777 (1935).

\bibitem{Bennett} C. H. Bennett, G. Brassard, C. Crepeau \textit{et al}.,
Phys. Rev. Lett. \textbf{70}, 1895 (1993).

\bibitem{Jozsa} A. Ekert and R. Jozsa, Philos. Trans. R. Soc. London \textbf{%
356,} 1769 (1998).

\bibitem{Bell} J. S. Bell, Physics \textbf{1}, 195 (1964).

\bibitem{Fu} J. Fu and S. Sun, arXiv:quant-ph/1003.1435v4.

\bibitem{Green} M. B. Green, J. H. Schwarz, and E. Witten, \textit{%
Superstring theory} (Cambridge University Press, Cambridge and New York,
1987).

\bibitem{Michio} K. Michio, \textit{Introduction to Superstring and M-Theory}%
, 2nd edn. (Springer-Verlag, New York, 1999).

\bibitem{Witten} E. Witten, Nucl. Phys. B \textbf{443}, 85 (1995).

\bibitem{Horava} P. Horava and E. Witten, Nucl. Phys. B \textbf{460}, 506
(1996); Nucl. Phys. B \textbf{475}, 94 (1996).

\bibitem{Dvali} G. Dvali, Ian I. Kogan, and M. Shifman, Phys. Rev. D \textbf{%
62}, 106001 (2000).

\bibitem{Furtado} C. Furtado, F. Moraes and V. B. Bezerra, Phys. Rev. D 
\textbf{59}, 107504 (1999); C. Furtado, V. B. Bezerra and F. Moraes,
Europhys. Lett. \textbf{52}, 1 (2000).

\bibitem{Capozziello} S. Capozziello, G. Lambiase1 and C. Stornaiolo,
Europhys. Lett. \textbf{48}, 482 (1999).

\bibitem{Kikkawa} K. Kikkawa, Phys. Lett. B \textbf{297,} 89 (1992); K.
Kikkawa and H. Tamura, arXiv:hep-th/9407160v1.

\bibitem{Berry} M.\thinspace V. Berry, Proc. R. Soc. A \textbf{392}, 45
(1984).

\bibitem{Shapere} A. Shapere, and F. Wilczek, \textit{Geometric phases in
physics} (World Scientific, Singapore, 1989).

\bibitem{Simon} B. Simon, Phys. Rev. Lett. \textbf{51}, 2167 (1983).

\bibitem{Aharonov} Y. Aharonov and D. Bohm, Phys. Rev. \textbf{115}, 485
(1959).

\bibitem{Zafiris} E. Zafiris, Journal of Mathematical Physics \textbf{47},
092103, (2006); E. Zafiris, "\textit{The Global Symmetry Group of Quantum
Spectral Beams and Geometric Phase Factors}", Advances Math. Phys. (to be
published).

\bibitem{Pedder} C. Pedder, J. Sonner, and D. Tong, Phys. Rev. D \textbf{77}%
, 025009 (2008).

\bibitem{Bredon} G. E. Bredon, \textit{Topology and Geometry}
(Springer-Verlag, New York, 1993); G. E. Bredon, \textit{Sheaf Theory}
(Springer-Verlag, New York, 1997).

\bibitem{Besicovitch} A. S. Besicovitch, \textit{Almost Periodic Functions}
(Cambridge University Press, Cambridge, 1932).

\bibitem{Corduneanu} C. Corduneanu, \textit{Almost Periodic Functions}
(Chelsea, New York, 1989).

\bibitem{Keh} K. Lii and M. Rosenblatt, The Annals of Statistics \textbf{34}%
, 1115 (2006).

\bibitem{Hong} Y. Han and J. Hong, J. Math. Anal. Appl. \textbf{336}, 962
(2007).

\bibitem{Spitters} B. Spitters, Logical Methods in Computer Science \textbf{1%
}, 1 (2005).

\bibitem{Dalfovo} F. Dalfovo, S. Giorgini, L. P. Pitaevskii, and S.
Stringari, Rev. Mod. Phys. \textbf{71}, 463 (1999).
\end{thebibliography}
\end{document}